# Imaging 3D polarization dynamics via deep learning 4D-STEM

Jinho Byun[1]†, Keeyong Lee[1]†, Myoungho Jeong[2], Eunha Lee[2], Jeongil Bang[3], Haeryong Kim [3],

Geun Ho Gu[1]*, Sang Ho Oh[1, 4]*

[1] Department of Energy Engineering, Korea Institute of Energy Technology (KENTECH), Naju, 58330, Korea

[2] Analytical Engineering Group, Samsung Advanced Institute of Technology, Suwon, 16678, Korea

[3] Thin Film TU, Samsung Advanced Institute of Technology, Suwon, 16678, Korea

[4] Center for Shared Research Facilities, Korea Institute of Energy Technology (KENTECH), Naju, 58330, Korea

* Corresponding author. Email: ggu@kentech.ac.kr (G.G.), shoh@kentech.ac.kr (S.H.O.)

† These authors contributed equally to this work.

**Recent advances in ferroelectrics highlight the role of three-dimensional (3D) polar entities in forming topological polar textures and generating giant electromechanical responses, during polarization rotation. However, current electron microscopy methods lack the depth resolution to resolve the polarization component along the electron beam direction, which restricts full characterization. Here, we present a deep learning framework combined with four-dimensional scanning transmission electron microscopy to reconstruct 3D polarization maps in $Ba_{0.5}Sr_{0.5}TiO_3$ thin-film capacitors with picometer-level accuracy under applied electric fields. Our approach enables observation of polar nanodomains consistent with the polar slush model and shows that switching occurs through coordinated vector rotation toward <111> energy minima, rather than magnitude changes. Furthermore, regions with higher topological density exhibit smaller polarization variation when the electric field changes, indicating topological protection. Our work reveals the value of 3D polarization mapping in elucidating complex nanoscale polar phenomena, with broad implications for emergent ferroelectrics.**

Accurate measurement of individual polarization vectors and their response to electric fields is essential for understanding the macroscopic ferroelectric properties of ferroelectric (FE) materials. This is especially critical for non-classical FEs, such as relaxor ferroelectrics (RFEs) and diffuse ferroelectrics (DFEs), in which polarization exhibits complex spatial distributions influenced by structural and chemical heterogeneities. These materials are characterized by polar nanodomains (PNDs), polarization rotation, low-angle domain boundaries, and intricate polar topologies[1,2,8]. Capturing such complex polarization phenomena requires resolving all three-dimensional (3D) components of the polarization vector with picometer-level accuracy and precision[10].

Atomic-resolution scanning transmission electron microscopy (STEM) imaging has been widely employed for mapping polarization at the unit-cell level in real space[2–6,11–13]. However, this technique inherently lacks sensitivity to polarization components oriented along the electron beam direction. When a full range of electrons scattered and diffracted are collected in reciprocal space, these signals can provide detailed information on the 3D atomic configuration via structure factor ($F_{\mathrm{hkl}}$), and on the spatial distribution of PNDs, via shape factor[14–16]. For instance, diffuse scattering signals around Bragg reflections in X-ray or neutron diffraction experiments yields global, mesoscale structural information about PNDs[15,16], albeit averaged across the probed volume[10,12,17,18]. Under kinematic diffraction conditions, the diffracted intensity ($I_{\mathrm{hkl}}$) is proportional to $|F_{\mathrm{hkl}}|^2$,

$$\boldsymbol{F}_{\mathbf{hkl}} = \sum_{\mathrm{j}} \boldsymbol{f}_{\mathrm{j}} \boldsymbol{e}^{-2\pi i(hx_{\mathrm{j}}+ky_{\mathrm{j}}+lz_{\mathrm{j}})} \tag{1}$$

where $x_j$, $y_j$, $z_j$ are the position of the $j$-th atom, and $f_j$ is the atomic scattering factor of the $j$-th atom.

Recent advancements in four-dimensional STEM (4D-STEM) have facilitated rapid acquisition of electron diffraction patterns (DPs) at each probe position, enabling powerful visualization of local polarization distributions. This technique has successfully revealed complex polar textures such as polar vortices, polar skyrmion, and polar meron[11,19–22]. However, extracting quantitative 3D polarization vectors from the highly complex and dynamically diffracted 4D-STEM datasets remains a significant challenge[19]. Previous approaches relying on the intensity asymmetry of Bijvoet pairs have offered predominantly qualitative insights into polarization[19]. Recent integration of machine learning methods with 4D-STEM analysis has begun to open new possibilities for quantitative polarization measurements[23–28]. Given the complexity of intensities in dynamical diffraction, influenced by sample thickness, tilt, and local chemical variations[29,30], a comprehensive interpretation of experimental 4D-STEM data demands supervised deep learning trained on simulated DPs spanning extensive parameter spaces.

To overcome these challenges and enable detailed characterization of PNDs, their topological states, and switching behaviors in FEs, we have developed an advanced deep learning framework specifically designed for 4D-STEM data analysis. Our model employs a Wasserstein autoencoder (WAE)[31] integrated with a predictor network to resolve all 3D components of the polarization vector —including the previously inaccessible electron-beam direction component—with picometer-level accuracy (mean absolute error < 1 pm). To bridge the gap between experimental and simulated datasets, we incorporated domain adaptation (DA)-based semi-supervised learning into our framework.

We demonstrate the capabilities of our method by investigating nanoscale polarization dynamics in $Ba_{0.5}Sr_{0.5}TiO_3$ (BST), a technologically significant RFE-like material used extensively in thin-film capacitors[4,32,33]. Although BST is not considered as a canonical RFE, it displays many phenomenological similarities to RFE, including diffuse phase transitions[34–36], nanoscale polar heterogeneities[4], and frequency-dependent dielectric properties[37]. Additionally, BST possesses high dielectric permittivity, excellent tunability, and relatively low dielectric loss[4,32–34,38], making it particularly attractive for dielectric energy storage applications[32,33]. Despite extensive studies, the fundamental mechanisms governing the polarization behavior of BST remain unclear. The complexity arises from the intrinsic diversity of polarization states, the presence of dynamic PNDs, and the fact that domain wall thickness is comparable to the size of domains.

Our analysis elucidates the intricate distribution of 3D polarization vectors and their response to applied electric fields, providing unprecedented insight into the diffusive character and rotational dynamics of polarization vectors in BST. The detailed analysis has been challenging for conventional techniques due to intrinsic limitations in capturing full 3D polarization. By resolving the 3D polarization vectors with high accuracy, our results reveal that the BST film comprises densely packed, interconnected PNDs that interact via continuous rotation of polarization vectors across domain boundaries. In-situ biasing experiments further show that polarization switching behavior closely correlates with the local topological polarization state, exhibiting coordinated adjustments in the $x$, $y$, and $z$ polarization components. These adjustments follow energetically favorable pathways, predominantly oriented along <111> directions towards the applied electric field, rather than resulting from substantial changes in polarization magnitude or extensive domain merging.

## Deep learning architecture

We first acquired experimental 4D-STEM datasets from an epitaxial BST thin film on a 63 × 165-pixel grid using an electron probe with a semi-convergence angle of 2 mrad, as illustrated in Fig. 1a. To facilitate supervised training of our model, we generated atomic structures of BST with matching composition and diverse 3D polarization states through molecular dynamics (MD) simulations (Fig. 1b). Multislice simulations were then performed using these structures under the same beam conditions as the experimental 4D-STEM, creating an extensive simulated dataset comprising 103,200 DPs (details in Methods). These DPs corresponded to 32,000 unit-cells with varying polarization states ($P_x$, $P_y$, $P_z$), sample tilt angles ($\theta_x$, $\theta_y$), and thickness ($t$), ensuring comprehensive coverage of experimental conditions (Fig. 1b).

Our WAE encoded DPs into a latent space, enabling a dense neural network to predict key parameters and reconstruct input DPs via a transposed convolution decoder (Fig. 1c). Initially, the model, trained exclusively on simulated DPs, showed only 51% overlap between experimental and simulated latent-space distributions (Extended Data Fig. 1a), highlighting discrepancies potentially leading to inaccurate predictions[39]. While DA methods have been conventionally used to align experimental and simulated domains[40–42], standard DA approaches risk significant performance degradation when ignoring intrinsic domain differences[43]. We identified the mismatch predominantly near diffraction disk regions, caused by factors unaccounted for in simulations, such as sample surface contamination, damage, inelastic scattering, and detector-related noise (e.g., shot noise, dark noise, readout noise). These factors significantly alter disk intensities in experimental DPs.

To address this, we implemented a semi-supervised learning strategy that incorporated both labeled simulated DPs and unlabeled experimental DPs within the WAE framework. Specifically, we developed a novel DA layer designed to effectively transform experimental DPs (Fig. 1d). This semi-supervised approach significantly enhanced domain alignment, increasing the overlap between experimental and simulated latent distributions to 91% (Extended Data Fig. 1b). Critically, the refined model accurately predicted independent parameters, as demonstrated comprehensively in Extended Data Fig. 2. This validation includes parity plots comparing predicted and ground truth values for thickness ($t$), sample tilt angles ($\theta_x$, $\theta_y$), and polarization components ($P_\mathrm{x}$, $P_\mathrm{y}$, $P_z$), achieving a mean absolute error (MAE) of less than 1 pm—well within typical polarization variations observed in BST[4]. Further validation of our DA method involved identifying simulated images closely matching experimental images using Euclidean distance metrics in parameter space (Extended Data Fig. 1c, d). These results confirmed that our approach substantially improved alignment between simulated and experimental domains and enhanced individual DP similarity, verifying the robustness and precision of our prediction framework.

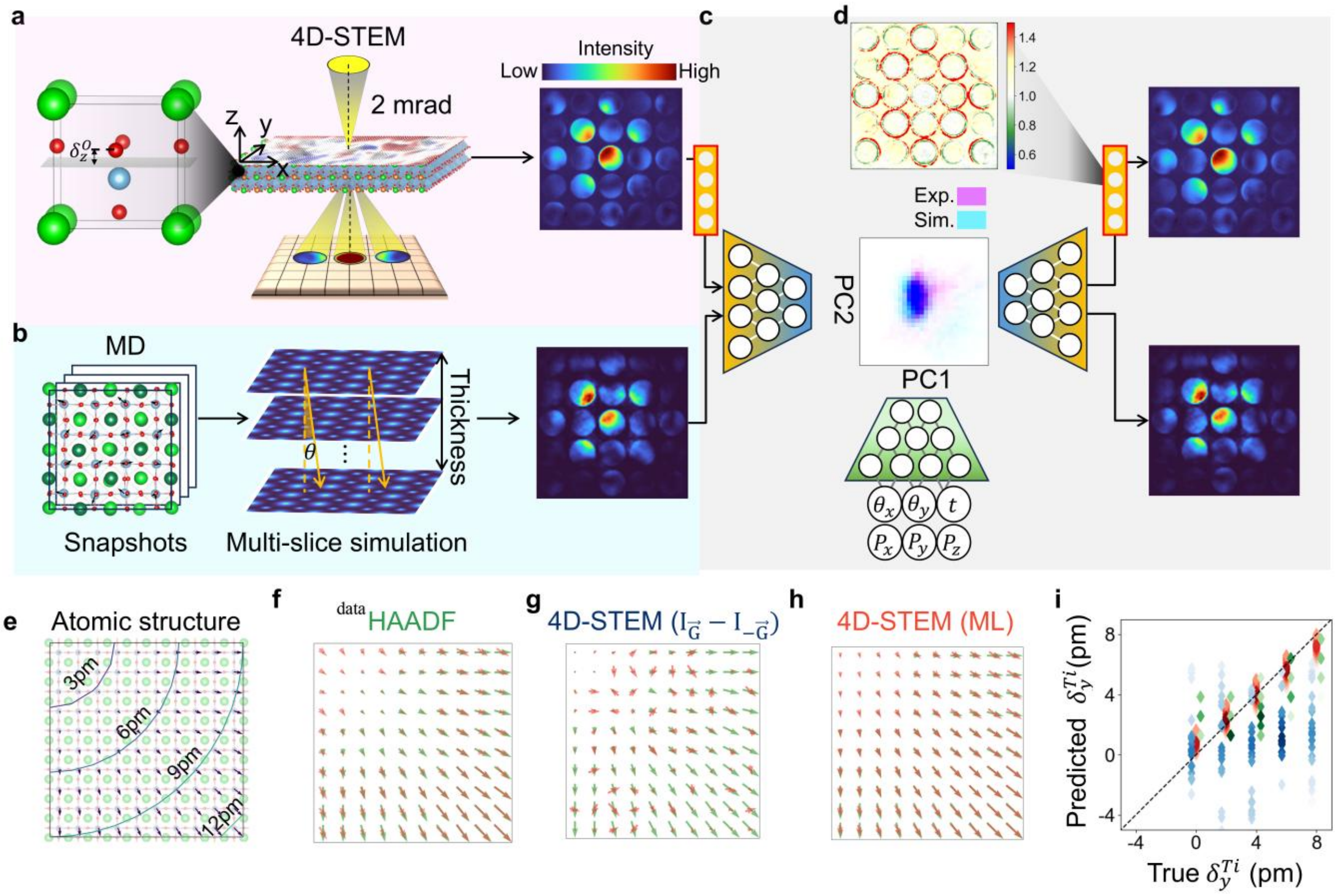

**Fig. 1 | Deep learning framework for determining picometer-scale displacement vectors from 4D-STEM data. a,** Schematics of the 4D-STEM experiment. Diffraction patterns (DPs) are collected by scanning an electron probe across a $Ba_{0.5}Sr_{0.5}TiO_3$ (BST) sample aligned along the [001] zone axis. **b,** Workflow of DP simulation. A range of atomic structures with various polarization states are generated by molecular dynamics (MD) simulation and used as inputs to multislice simulations (see Methods). The simulated dataset spans diverse 3D polarization ($P_x$, $P_y$, $P_z$), tilt angles ($\theta_x$, $\theta_y$), and thickness ($t$), covering the parameter space of the experiment. **c,** Network architecture. A Wasserstein autoencoder (WAE) encodes DPs into a latent space, and a decoder reconstructs DPs from latent space. A multilayer perceptron (MLP) predicts the underlying properties. To address discrepancies between simulated and experimental DPs, a domain adaptation (DA) layer transforms experimental DP to closely match the simulated DP, ensuring consistent encoding by the same WAE encoder. **d,** Example transformation induced by the DA layer (weights shown), aligning experimental and simulated DPs. **e,** BST model structure with a continuous variation of displacement from top-left to bottom-right. **f-h**, Displacement maps predicted by atomic-column position picking in high-angle annular dark-field (HAADF) STEM image (**f**), conventional intensity asymmetry analysis using the difference between diffraction disk intensities ($I_{\mathrm{hkl}} - I_{\overline{\mathrm{hkl}}}$) in 4D-STEM (**g**), and the present deep-learning approach (**h**). Green arrows denote ground truth; red arrows are predictions. **i,** Parity plots comparing ground-truth vs. predicted displacement for HAADF (green), intensity asymmetry (blue), and deep-learning (red) of 4D-STEM data. The significantly improved agreement validates the deep-learning analysis.

To evaluate the model's ability to capture polarization variations within PNDs, we analyzed simulated 4D-STEM data from a polar system featuring a gradient in Ti displacement ($\delta_{\mathrm{Ti}}$) from the top-left to the bottom-right and random Ba/Sr distributions at the A-site (Fig. 1e). As a comparison, direct $\delta_{Ti}$ measurements from simulated STEM Z-contrast images[44] showed considerable noise distribution for the small $\delta_{Ti}$s, limiting its reliability (Fig. 1f). Conventional 4D-STEM methods, based on breakdown of Friedel's pair symmetry (intensity asymmetry between diffraction disks $I_{\mathrm{hkl}}$ and $I_{\overline{\mathrm{hkl}}}$) [45–48], often incorrectly reflect local Ba/Sr arrangements rather than true $\delta_{Ti}$ due to the probe's nanoscale size[30] (Fig. 1g). In contrast, our machine learning model accurately predicts the full range of $\delta_{Ti}$ with remarkable consistency (Fig. 1h). Figure 1i summarizes the quantitative agreement between predicted and true $\delta_{Ti}$ values, demonstrating the superior capability of our deep learning framework for 4D-STEM analysis compared to conventional intensity measurement.

## 3D polarization mapping via deep learning of 4D-STEM

To elucidate the capability of our model in predicting the polarization component along the electron beam direction ($P_z$), as well as the in-plane polarization components ($P_x$, $P_y$), we analyzed the derivatives of DP intensities with respect to $t$, $\theta_y$, $P_y$ and $P_z$ (Extended Data Fig. 3). The derivatives with respect to each parameter are distinctly recognizable, clearly demonstrating the model's reliability and independent detection capability. Furthermore, even when different combinations of parameters were tested simultaneously, each intensity derivative remained highly distinguishable.

Given the relatively small intensity variations in DPs induced by changes in $P_z$, we evaluated the practical detectability of these subtle changes by examining the MAE of polarization predictions under varying levels of DP noise (Extended Data Fig. 4). At lower electron doses, the MAE became significantly large, exceeding typical polarization magnitudes observed in BST[4], indicating potential limitations under low-dose conditions. However, at electron doses typical of our experimental conditions (~$10^6$ electrons per DP), the MAE remained substantially smaller than the characteristic polarization magnitude of BST, affirming the robustness of the polarization predictions under realistic experimental conditions.

The latent-space alignment achieved during predictor training provided valuable insights into DP interpretation[49]. To investigate this further, we calculated the Pearson correlation matrix between the predicted properties—such as $t$, $\theta_x$, $\theta_y$, $P_x$, $P_y$, and $P_z$—and the latent vector features ($L$i, i = 1-24) (Extended Data Fig. 5a). The correlation analysis identified specific latent vectors associated with particular physical parameters: $L2$ and $L10$ strongly correlated with $P_y$, while $L17$ and $L19$ were primarily associated with $P_z$, highlighting the model's ability to disentangle complex DP data into independent latent-space dimensions. Extended Data Fig. 5b, c further visualizes this disentanglement through scatter plots of data projected onto 2D latent spaces, color-coded by physical parameters ($t$, $\theta_y$, $P_y$, $P_z$). Notably, $P_y$ shows clear correlation within the $L2/L10$ latent subspace (Extended Data Fig. 5b), and $P_z$ demonstrates strong correlation within the $L17/L19$ latent subspace (Extended Data Fig. 5c). These results confirm that our deep learning approach effectively deconvolutes intricate interplays among sample properties encoded in DPs, significantly enhancing the interpretative power of 4D-STEM analysis.

## 3D polarization map revealing polar nanodomains and their topologies

We applied our deep learning model to a 20 nm-thick epitaxial BST film sandwiched between top and bottom $SrRuO_3$ (SRO) electrodes (Fig. 2a, b). As shown in Fig. 2b, the BST epitaxial film was structurally defect-free and established coherent interfaces with the SRO electrodes. The thickness of the TEM sample determined by our deep learning model—an important parameter significantly influencing DP intensities—shows excellent agreement with the thickness map obtained by electron energy loss spectroscopy (EELS), as presented in Extended Data Fig. 6, thereby validating our approach.

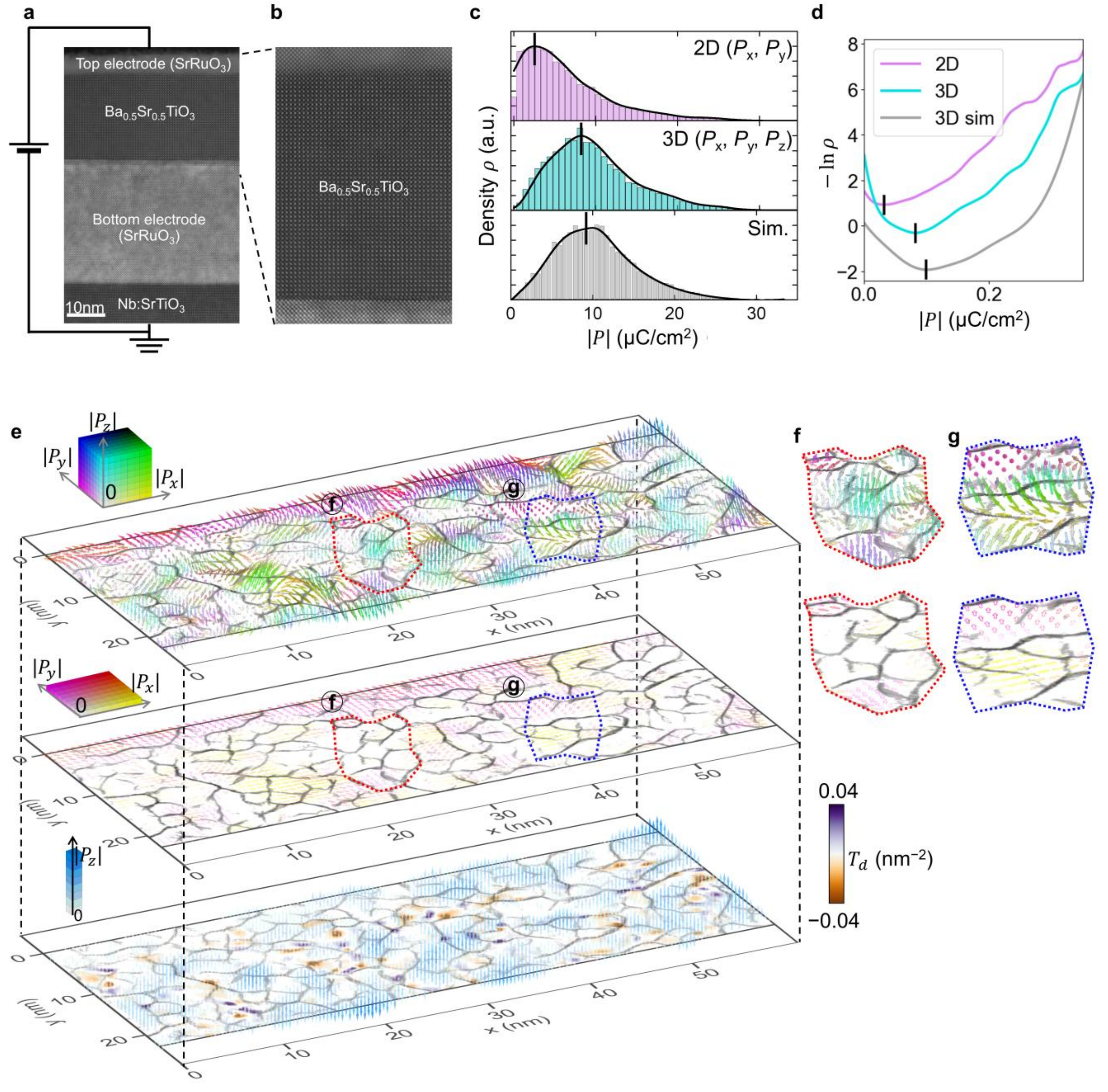

**Fig. 2 | Three-dimensional polarization state of epitaxial BST film revealed by 4D-STEM analysis. a,** Cross-sectional HAADF-STEM image of the epitaxial BST thin film on a Nb:$SrTiO_3$ substrate, with $SrRuO_3$ (SRO) as top and bottom electrodes. **b,** Magnified HAADF-STEM image showing the defect-free BST layer and coherent interfaces with the SRO electrodes. **c,** Probability density ($\rho$) plots of the polarization magnitude for (top) 2D vector components (projected in the *xy*-

plane), (middle) the full 3D vector components, and (bottom) a MD simulation for comparison. **d**, Free-energy-like potential obtained by taking the negative logarithm of the density in **c**. **e,** Maps of the BST polarization represented as 3D vectors (top), 2D *xy*-projections (middle), and the 1D *z*-component (bottom). Gray lines denote domain boundaries, identified where polarization changes are locally maximized (see Methods). The bottom panel overlays $|P_z|$ on a topological density ($T_d$) map, highlighting their spatial correlation. **f,g,** Enlarged views illustrating polarization rotation across domains (**f**) and at domain walls (**g**), with full 3D vectors (top) and 2D *xy*-projections (bottom). Vectors dominated by the *z*-component can appear nonpolar if only the *xy*-plane is considered, underscoring the importance of full 3D analysis.

Figure 2c presents the density plots of the polarization magnitude measured experimentally and those derived from MD. Notably, significant differences emerged between plots utilizing the full 3D polarization vectors and those relying solely on 2D in-plane components (projected onto the *xy*-plane). The 3D measurement closely matched the MD simulation results, highlighting the importance of accurately capturing all three polarization components ($P_x$, $P_y$, $P_z$). Given our TEM sample geometry, the $P_z$ component equivalently corresponds to the in-plane $P_x$ component, reinforcing the critical role of the $P_z$ component in determining both polarization magnitude and orientation. Figure 2d presents free-energy-like potential curves derived from the polarization magnitude distributions. The experimental 3D measurement and MD simulation reveal minima at finite polarization magnitudes of approximately 8 $\mu C/cm^2$ and 10 $\mu C/cm^2$, respectively. In contrast, the 2D measurements (*xy*-plane only) exhibited a significantly lower magnitude minimum (~3 $\mu C/cm^2$), underscoring the necessity of including the full 3D vector components for accurate characterization.

The polarization map incorporating all 3D polarization components ($P_x$, $P_y$, $P_z$) is presented in Fig. 2e, alongside a map constructed using only the 2D in-plane components ($P_x$, $P_y$) for comparison. A prominent observed feature is the coordinated rotation of polarization vectors, forming complex, nanoscale topological patterns. By defining domain boundaries as regions of maximum polarization variation (Extended Data Fig. 7), we identified nanoscale domains typically spanning 2-16 nm (mean size ~9 nm). Polarization vectors continuously rotate across these boundaries, with varying degrees of angular differences between adjacent domains (Extended Data Fig. 7).

Importantly, the conventional polarization map using only 2D components can incorrectly suggest nonpolar regions where the $P_z$ component dominates, as well as misleadingly identify certain regions dominated by $P_x$ and/or $P_y$ components as polar nanoregions (PNRs) (Fig. 2f, g)[4,50]. In contrast, our 3D polarization map reveals these apparent nonpolar areas as genuine nanodomains (Fig. 2f). Hence, the 3D polarization aligns more closely with the PND model described by the polar slush model, rather than the simpler PNR model. To validate this observation, we selected regions where polarization changes are dominant and variations from tilt or thickness are minimal (Extended Data Fig. 7d-g). We found that polarization fluctuations between neighboring nanodomains—particularly including those involving $P_z$ changes—are distinctly represented by intensity variations in experimental DPs.

The comprehensive 3D polarization analysis further enabled mapping topological charge ($Q$) and topological density ($T_d$), defined as:

$$Q = \int T_d(x, y)dxdy, \quad (2)$$

$$T_d\,(x, y) = \frac{1}{4\pi}\vec{\boldsymbol{u}} \cdot \left(\partial_x \vec{\boldsymbol{u}} \times \partial_y \vec{\boldsymbol{u}}\right), \quad (3)$$

where $\vec{\boldsymbol{u}}(x, y)$ is the normalized 3D net polarization vector on the *xy*-plane. The resulting topological density map (third layer in Fig. 2e) reveals the spatial distribution of local polar textures critical to polarization switching dynamics. Lower topological densities were observed near the top and bottom electrodes, regions dominated by low-angle domain boundaries. Conversely, higher topological densities were found centrally within the BST film, predominantly localized along high-angle domain boundaries, with distinct maxima typically appearing at cusps along these boundaries.

**Polarization dynamics under electric fields**

We investigated the switching behavior of PNDs within the BST film by analyzing 4D-STEM datasets obtained during in-situ electrical biasing experiments. External bias was applied to the top SRO electrode while the bottom SRO electrode and Nb:$SrTiO_3$ (Nb:STO) substrate were grounded (Fig. 2a). Initially, we examined the polarization configuration along the field direction ($P_y$) and its response under varying electric fields (Fig. 3). Near the top electrode, $P_y$ points upward, whereas it points downward near the bottom electrode, forming a characteristic tail-to-tail polarization configuration. High topological density is concentrated around the mid-region of the BST film, coinciding with weakened $P_y$ magnitudes and the convergence of opposing polarization directions. This tail-to-tail arrangement arises from symmetric lattice constrains at the epitaxial interfaces imposed by the SRO electrodes (Fig. 3b, e). To verify the influence of epitaxial constraints on polarization, we conducted MD simulations with constrained and relaxed BST cells (Extended Data Fig. 8). The constrained cell, with top and bottom surfaces fixed to STO to mimic epitaxial constraints, displayed a clear tail-to-tail $P_y$ configuration, closely matching experimental observations despite minor asymmetries. In contrast, the fully relaxed BST cell exhibited no preferred orientation, confirming that symmetric compressive strain promotes the tail-to-tail polarization arrangement.

Application of external bias influenced the polarization configuration significantly. Under -2 V bias, upward-pointing $P_y$ polarization (red arrows) expanded toward the lower half of the BST film, accompanied by reduced topological density (Fig. 3a, d). Conversely, under +2 V bias, downward $P_y$ polarization (blue arrows) grew toward the upper half (Fig. 3c, f). Areas with high topological density showed smaller polarization changes (orange arrows, Fig. 3g–i). Analysis of polarization variation and average topological density profiles under bias confirmed a clear inverse relationship between polarization responsiveness and local topological density (Fig. 3j-l), suggesting that topologically non-trivial states are more resistant to external bias-driven polarization changes, analogous to topological protection observed in skyrmions[51]. Extended Data Fig. 9 compares changes in local polarization components and tilts under external bias (–2 V, 0 V, and +2 V). The results indicate that most of the observed local fluctuations in DP intensities under external biases arise from polarization changes, whereas overall tilt variations are marginal.

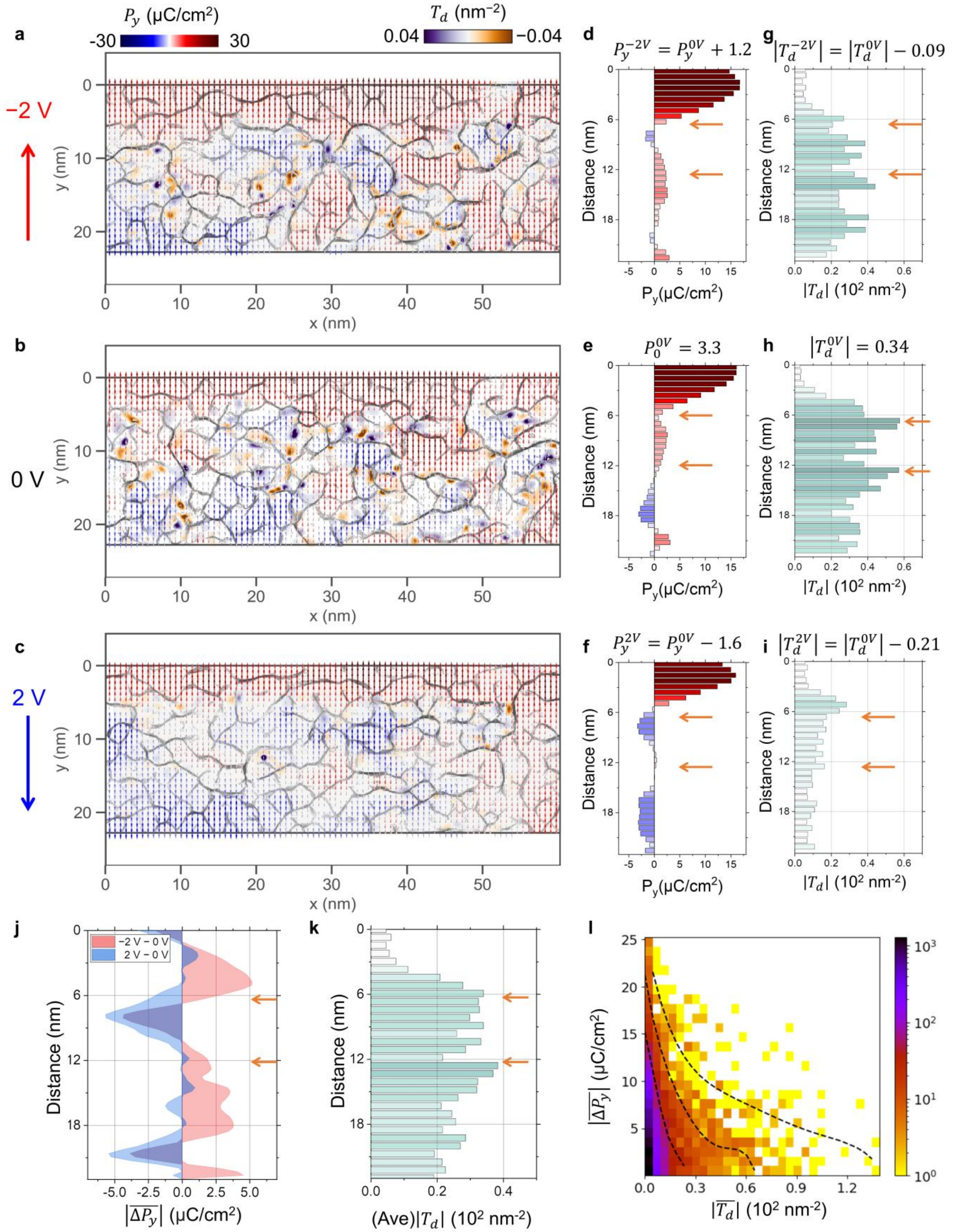

**Fig. 3 | Switching of polarization vectors towards the field direction mediated by topological state. a-c,** Maps of the polarization vector component projected onto the y-axis ($P_y$) at -2 V (**a**), 0 V

(**b**), and 2 V (**c**). The topological density ($T_d$) and domain boundaries (gray lines) are superimposed. The sign and magnitude of $P_y$ and $T_d$ are defined on each color bar. **d-f,** In-plane averaged vertical profiles of $P_y$ corresponding to the three bias conditions in **a-c**. **g-i,** In-plane averaged vertical profiles of $T_d$; orange arrows mark regions of pronounced topological density. **j,** In-plane averaged vertical profiles of the polarization difference ($\Delta P_y$) between -2 V to 0 V (red), and between 2 V and 0 V (blue). **k,** In-plane averaged vertical profiles of $T_d$ averaged over the three external bias conditions shown in **g-i**. **l,** 2D histogram correlating the average magnitude of change in $P_y$, $|\overline{\Delta P_y}|$, with the average local topological density, $|\overline{T_d}|$. The color scale indicates the number of states per pixel (log scale). Dashed curves, obtained by fitting bins with similar populations, reveal an inverse correlation between $|\overline{\Delta P_y}|$ and $|\overline{T_d}|$.

Notably, electric fields influenced not only the $P_y$ component aligned with the field but also the perpendicular components ($P_x$ and $P_z$) (Fig. 4a, Extended Data Fig. 10a-c). Consequently, all 3D vector components must be considered when describing polarization behavior under electric fields. Analysis of each component reveals that the 180° switching of the $P_y$ component is accompanied by coordinated changes in both the $P_x$ and $P_z$ components (Fig. 4c, d), resulting in the rotation of polarization vectors towards the direction of the electric field. This rotation modifies the angular range of domain misorientation (Fig. 4e), narrowing the initial broad distribution (full width at half maximum, FWHM of ~53°) to more focused distributions under ±2 V bias, such as FWHM of 35° at –2 V and that of 30° at +2 V. During this rotation, neither the net polarization magnitude $|P|$ nor the average domain size showed significant changes (Extended Data Fig. 10d, e), emphasizing that polarization vector rotation, rather than magnitude alteration or domain merging, is the primary response mechanism. This dominant rotational response is analogous to a Goldstone mode in 3D version of "Mexican hat" potential energy surface (PES), where polarization reorients along a continuous minimum-energy brim, unlike simpler 1D Landau models that favor magnitude changes[52].

The angular distribution and rotation trajectories of polarization vectors under applied electric fields can be effectively visualized by representing the vectors as poles in a stereographic projection (SP) (Fig. 4b). A reference sphere was constructed with a radius equal to the average polarization magnitude $|P|$, and the polarization vector components ($P_x$, $P_y$, $P_z$) were converted into spherical coordinates. Prior to applying an electric field, poles are aggregated into two primary groups against a diffuse background. To differentiate the depth of origin from the top to bottom electrode, poles are color-coded: red for the upper region (located predominantly in the northern hemisphere), blue for the lower region (situated slightly below the equator), corresponding to the tail-to-tail configuration in the $P_y$ map, and green for the central region of the BST film, which is dispersed as a diffuse background. Under an upward field (–2 V), the blue poles rotate toward the northern hemisphere, leaving traces along a great circle, while the green poles congregate and merge into the blue poles. Conversely, under a downward field (+2 V), the blue poles rotate toward the southern hemisphere along a similar great circle, again with green poles merging into them, but moving in the opposite direction. As poles become densely populated within each group and align more closely with each other, the curl defining the topological state of the polarization vectors decreases. This rotational alignment, which reduces the misorientation between neighboring domains under applied fields, promotes the aggregation of PNDs into larger clusters, particularly middle and bottom electrode regions. Nevertheless, most domains remain at the nanoscale, and the overall tail-to-tail polarization configuration persists.

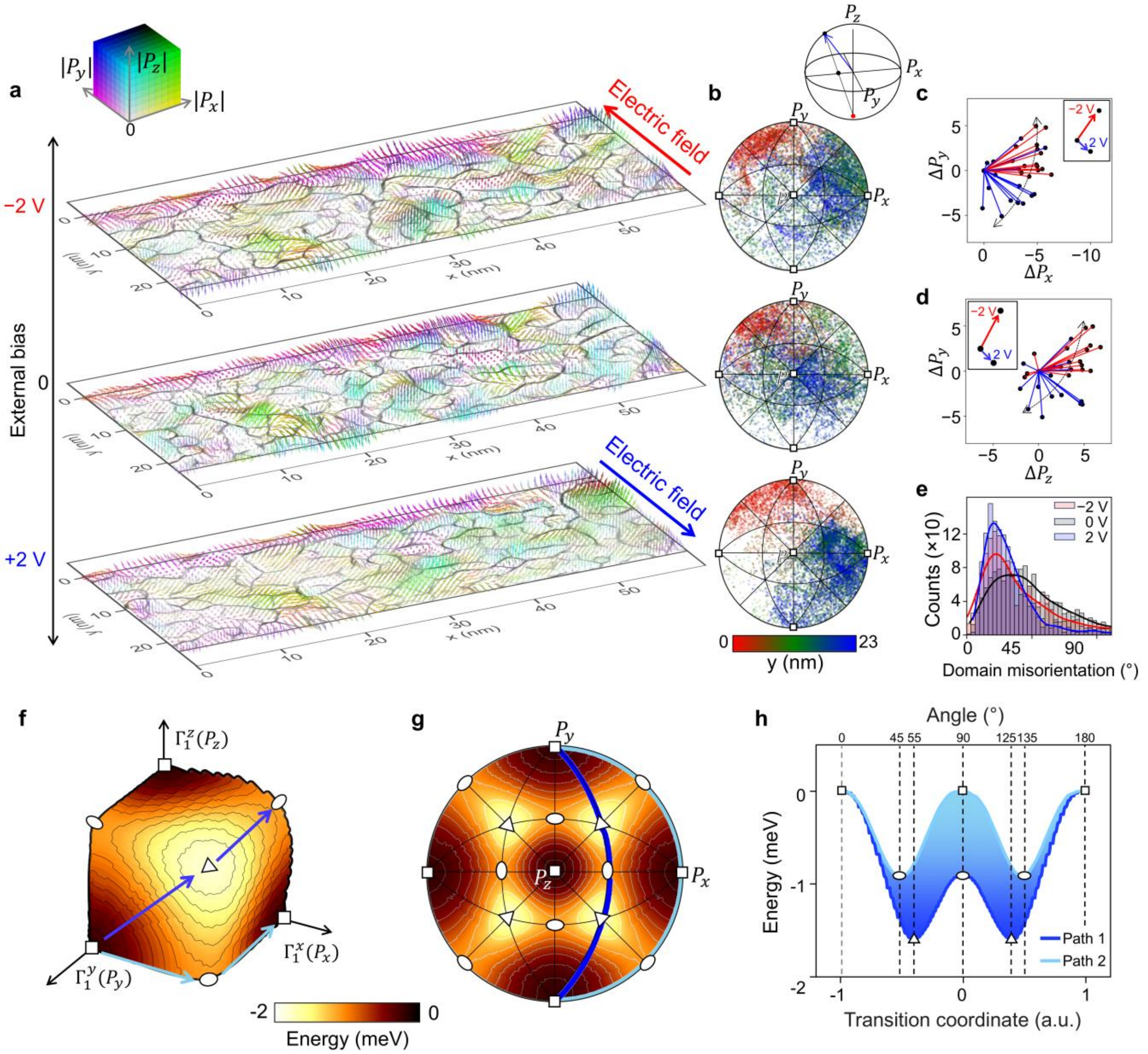

**Fig. 4 | 3D polarization analysis under electric fields reveals rotation of polarization vectors toward stable orientations. a**, 3D polarization vector field maps measured at -2 V, 0 V, and +2 V, showing how polarization vectors behave and domain structure changes under electric fields. **b,** Stereographic projections of the 3D polarization vectors, with poles color-coded by their vertical position ($y$, in nm) in the BST film, highlighting the evolution of the tail-to-tail domain configuration. Under an upward electric field (–2 V), the blue poles rotate toward the northern hemisphere, leaving traces along a great circle, while the green poles congregate and merge into the blue poles. An opposite behavior of poles was observed under a downward field (+2 V). **c,d,** Scatter plots comparing changes in polarization components $\Delta P_x$ (**d**), $\Delta P_z$ (**e**) with $\Delta P_y$, for three biasing conditions. The red line denoted switching from 0 V to -2 V, and the blue lines denote switching from 0 V to +2 V. Insets show the average polarization vector change during each switching process, indicating coordinated

reorientation. **e,** Histograms of the domain rotation angle for -2 V (red), 0 V (black), and +2 V (blue), illustrating field-induced alignment. **f,** Calculated 3D potential energy surface (PES) as a function of distortion modes corresponding to polarization components ($\Gamma_1^x(P_x)$, $\Gamma_1^y(P_y)$, $\Gamma_1^z(P_z)$). Local minima of the PES identify stable polarization states. The blue and light-blue lines represent two minimum-energy paths (MEPs): path 1 (blue) passing through (*a*, *a*, *a*) order parameter, and path 2 (light blue) constrained in the *xy*-plane. Symbols indicate the specific stable poles ({100}, {110}, {111}). **g,** Stereographic projection of the PES in **f**, showing 2D energy contours and two MEPs connecting low-energy {111} and {110} poles. **h,** Calculated energy profiles along path 1 (blue) and path 2 (light blue), connecting opposite poles of [0$\bar{1}$0] and [010] poles as a function of generalized transition coordinate. Shading area highlights the intermediate transition region between path1 and path 2.

The observed polarization dynamics were further validated by theoretical calculations of the PES for three degenerated distortion modes ($\Gamma_1^x(P_x)$, $\Gamma_1^y(P_y)$, $\Gamma_1^z(P_z)$) in BST. Calculations identified energy minima at the (*a*, *a*, *a*) and (*a*, *a*, 0) displacement modes, with the (*a*, *a*, *a*) configuration corresponding to the lowest-energy state (Fig. 4f), characteristic of RFEs where polarization along the <111> direction minimizes local lattice distortions[1,9,10]. Minimal energy paths (MEPs) connecting these minima revealed preferred polarization rotation pathways aligned with low-energy <111> (dark blue) and <110> orientations (light blue) (Fig. 4f).

The PES visualized using SP reveals low-energy potential wells at the {111} and {110} poles, corresponding to stable polarization orientations (Fig. 4g). MEPs connecting these minima are identified. Notably, the MEP is observed to lie exactly along the ($\bar{1}$01) trace (marked by a blue line in Fig. 4g), linking energy minima at adjacent {111} poles while the second MEP constrained on *xy*-plane naturally on (001) trace. Energy profiles along the two distinct MEPs exhibit double potential wells having two local minima at {111} and {110} poles (Fig. 4h).

When an electric field is applied along the *y*-direction (corresponding to the [010] or [0$\bar{1}$0]), polarization vectors rotate towards the nearest potential well aligning most closely with the applied field direction. Consequently, the polarization vector poles aggregate into these potential wells, displaying distinct clustering and trajectory patterns in experiments. Experimental polarization vectors clustered near these predicted {111} energy minima but remained dispersed around them, indicative of local topological constraints imposed by the existing PNDs structure hindering perfect alignment. Furthermore, polarization vectors do not align perfectly parallel to the field direction; instead, they settle into the stable orientation provided by the {111} potential well. These stable orientations at the {111} poles are achieved through coordinated modulation of all polarization components ($P_x$, $P_y$, $P_z$), allowing them to have magnitude of all polarization components (Extended Data Fig. 10a-c).

In summary, this study provides a comprehensive analysis of polarization dynamics in BST under applied electric fields, clearly demonstrating that polarization switching predominantly involves vector rotation towards energetically favorable orientations rather than alterations in polarization magnitude or substantial domain merging. By employing advanced deep learning methodologies to interpret 4D-STEM diffraction patterns, we successfully resolved all 3D polarization components, including those previously inaccessible through conventional methods. Our results emphasize the crucial influence of local topological states and the nanoscale nature of polarization domains on the switching behavior of polarization vectors. The insights gained from this approach significantly enhance our understanding of complex polarization phenomena in ferroelectric materials, paving the way for improved device functionalities and more accurate modeling of polarization behaviors in RFEs and related materials.

## Methods

**Thin film deposition.** A 20 nm-thick $Ba_{0.5}Sr_{0.5}TiO_3$ (BST) thin-film capacitors were grown by pulsed laser deposition (PLD) using a KrF excimer laser (λ = 248 nm). Before growth, a Nb-doped $SrTiO_3$ (100) substrate was etched in a buffered hydrofluoric acid etchant and annealed at 1000 °C for 1 h to achieve a $TiO_2$-terminated surface. The $SrRuO_3$ electrodes and BST layer were grown in an oxygen atmosphere of 100 mTorr, while the substrate temperature was maintained at 700 °C. A laser repetition rate of 10 Hz was used for depositing $SrRuO_3$ bottom electrode, the BST dielectric layer, and $SrRuO_3$ top electrode sequentially.

**TEM sample preparation.** The cross-sectional samples used for in-situ STEM biasing experiments were prepared on a MEMS chip (FIB-Optimized, Protochips) via $Ga^+$ ion beam using a focus-ion-beam (FIB, Helios 5 UX, Thermofisher Scientific). After initial thinning at 30 kV, the ion beam was reduced to 5 kV to minimize damage from ion milling.

**HAADF STEM imaging.** The high-angle annular dark-field (HAADF) imaging used a detection angle in the range of 62-200 mrad. The convergence semi-angle of the focused probe was 21.4 mrad. The scanning distortion was corrected by comparing the image with a 90-degree rotated image[53]. The information of each image's fast scanning direction was used as a reference for correcting the scanning distortion of the slow scanning direction.

**Thickness measurement using STEM EELS.** EELS 2D scan data was acquired by a Gatan K3 camera with a 5 mm diameter EELS aperture. The data was recorded in the energy range of 0-2900 eV with dispersion of 0.9 eV. The EELS dataset was acquired with a scanning step of 0.3 nm and 254 × 120 probe positions. The data was acquired with a dwell time of 20 ms per frame. The t/λ map was obtained by EELS log-ratio method at region of the BST where 4D-STEM data was acquired. The inelastic mean free path for BST was calculated as 111.7 nm by Malis model with collection angle of 30 mrad and acceleration voltage of 300 kV[54].

**4D STEM data acquisition.** In-situ 4D-STEM measurements under applied bias were conducted in an aberration-corrected STEM (Spectra Ultra, Thermofisher Scientific) at 300 kV with a 2.2 mrad probe convergence semi-angle. The 4D-STEM data were acquired using a Gatan K3 camera attached to the energy filter (Gatan Quantum HR) that removed electrons with energy losses above 5 eV. Each diffraction pattern (DP) was collected as a 512 × 512 pixel image with a dwell time of 7 ms per frame. A scanning step of 0.4 nm was chosen, resulting in 180 × 110 probe positions. Initial DPs were denoised using a tensor singular value decomposition approach[55].

**Generation of atomic models.** Machine learning molecular dynamics (MLMD)[56,57] was performed as implemented in VASP to simulate all the possible polarization states of BST. Machine Learning Force Field (MLFF) was trained in an isothermal-isobaric (NpT) ensemble with a time step of 1 fs and a temperature of 1000 K, using the 2 × 2 × 2 supercells of BST with checkerboard-like Ba/Sr ordering. Based on the obtained MLFF, 10 ps canonical (NVT)-MLMD simulations were run for 8 × 8 × 2 supercells of BST. The temperature was controlled at 300 K using the Nose–Hoover thermostat. Additionally, a diverging polar system with Ti displacement ($\delta_{Ti}$) gradually increasing by 2 pm is considered. For these structures, each initial A-site cation position was randomly assigned as Sr or Ba with equal (50%) probability.

**4D STEM simulations.** DPs were simulated from MD snapshots of BST using the multislice method, as implemented in the abTEM[58] package. These simulations were conducted on 516 unique Titanium (Ti) columns, excluding duplicate sites based on a polarization tolerance of 1.5 pm. The sample thickness was varied from 30 to 80 nm in 1 nm increments. To incorporate small angular deviations of the transmitted beam from the [001] direction, beam tilts ranging from -2 mrad to +2 mrad were introduced by modifying the propagator function. The convergence semi-angle was set to 2.2 mrad, and the acceleration voltage was maintained at 300 kV. This comprehensive parameter sweep resulted in the generation of 103,200 DPs.

**Model architecture.** Our model integrates a Wasserstein autoencoder (WAE)[59] with a predictor network to simultaneously reconstruct diffraction patterns and extract physical parameters from 4D-STEM data. In the encoder, a series of six convolutional blocks reduces the spatial dimensions while increasing the feature depth and finally into a 2048 feature vectors. Each block consists of a convolutional layer, batch normalization, and a Swish activation. These feature vectors are further compressed into a 32-dimensional latent vector through a fully connected layer (FCL). For reconstruction, the decoder takes the latent vector and passes it through a FCL to expand it back into a 2048 feature vectors, followed by a series of transposed convolutional layers that gradually recover the spatial resolution of the diffraction pattern. Parallel to the reconstruction pathway, a predictor network is employed to extract key physical parameters such as tilts ($\theta_x$, $\theta_y$), thickness ($t$), displacements ($\delta_{\mathrm{Ti}}$, $\delta_{\mathrm{O}}$), and three polarization components ($P_{\mathrm{x}}$, $P_{\mathrm{y}}$, $P_{\mathrm{z}}$) from 24 latent vectors. This network uses two sequential layers, each followed by FCL, layer normalization, and Swish activation function, to map the latent representation to the desired output parameters. For experimental DPs, the DA layers were introduced before and after the WAE to adjust the intensity at each pixel.

**Model Training.** The integrated WAE-predictor model was trained by minimizing a composite loss function:

$$L_{total} = \lambda_{recon}L_{recon} + \lambda_{MMD}L_{MMD} + \lambda_{pred}L_{pred} \tag{3}$$

where $L_{\mathrm{recon}}$ represents the loss for diffraction pattern reconstruction, $L_{\mathrm{MMD}}$ is the latent space regularization calculated via Maximum Mean Discrepancy (MMD) between the latent representations of simulated and experimental data with respect to a Gaussian distribution, and $L_{\mathrm{pred}}$ is the loss for physical parameter prediction (using Mean Squared Error, MSE). The specific weights were set to ($\lambda_{\mathrm{recon}}$, $\lambda_{\mathrm{MMD}}$, $\lambda_{\mathrm{pred}}$) = (1, 100, 5). Optimization was performed using the Adam optimizer with a learning rate of $1 \times 10^{-3}$. The model was trained for up to 8 epochs with a batch size of 32, implementing an early stopping mechanism triggered by MMD loss divergence on the validation set. Both input simulated 4D-STEM diffraction patterns and target physical parameters (tilt, thickness, displacement, polarization) were standardized.

**Domain boundary identification and domain wall measurement.** Domain boundaries were identified as regions exhibiting a maximum change in local polarization among the neighboring sites[6]. To quantify local polarization variation, the variance for each polarization component ($P_x$, $P_y$, $P_z$) was first calculated within a 3 × 3 unit-cell window. The overall 3D polarization standard deviation ($\sigma_P$) was considered as the square root of variance of each component. The domain boundaries were identified from the $\sigma_P$ map using an iterative morphological skeletonization algorithm performed for 10 iterations. In each iteration, the current map was eroded. This eroded map was then subjected to a morphological opening operation. The difference between the eroded map and its opened map was accumulated onto the skeleton map via a pixel-wise maximum operation. After locating the domain boundaries, the $\sigma_P$ was mapped onto the domain boundaries (Extended Data Fig. 7). The domain misorientation angles were measured at domain boundaries by measuring the angle between two representative vectors (Fig. 4e). The domain size was measured from one domain boundary to the next domain boundary along the *x* and *y* directions. This process was repeated for each x-pixel and y-pixel in one-dimensional slices, and the results were plotted as a histogram (Extended Data Fig. 10).

**Potential energy surface calculation.** The potential energy surface (PES) was generated by perturbing the high-symmetry cubic structure along eigen displacements of lowest soft-mode at Γ points derived from its force constant matrix[60]. These eigen displacements can form triply degenerate sets of orthogonal vectors. The three-dimensional (3D) distortion space (Fig. 4f) was spanned by three orthogonal basis displacements, labeled as $\Gamma_1^x$, $\Gamma_1^y$, and $\Gamma_1^z$. The energy landscape within this 3D space was mapped by distorting the crystal structure along linear combinations of these basis eigenvectors. The minimum energies were mapped onto the PES. The minimal energy path (MEP, blue curve in Fig. 4g) was obtained via gradient descent starting from the local minima located at the poles of the $\{010\}$ and $\{101\}$ orientations.

**Data availability**

The datasets generated during and/or analyzed during the current study are available from the corresponding authors on reasonable request. The machine learning model will be available at Github repository after the acceptance.

**Acknowledgments** This research was supported by the Samsung Advanced Institute of Technology (SAIT), Samsung Electronics Co., Ltd (IO241114-11192-01), and the Technology Innovation Program (RS-2024-00418991, Development of In situ/Operando Analysis Technology for Package Reliability and Failure Identification) funded by the Ministry of Trade, Industry & Energy (MOTIE, Korea). The TEM work at Korea Institute of Energy Technology (KENTECH) was supported by the KENTECH Research Grant (KRG2022-01-019), and KENTECH Center for Shared Research Facilities. G.G. acknowledges support from the Open R&D program of Korea Electric Power Corporation (No.R23XO03), the generous supercomputing time from KISTI, and the National Research Foundation of Korea (NRF) grants funded by the Korean government (Ministry of Science and ICT, MSIT) (RS-2024-00449449).

**Author contributions** S.H.O. and G. G. conceived the project; J.B. conducted the 4D-STEM analysis and deep learning implementation under the supervision of S.H.O. and G.G.; K.L., M.J. and E.L. performed TEM experiment; J.I.B. and H.K. conducted thin film growth; all authors contributed to the interpretation of data and visualization of results; J.B. and S.H.O. prepared the paper, which was reviewed and edited by all authors

**Competing interests:** The authors declare no competing interests.

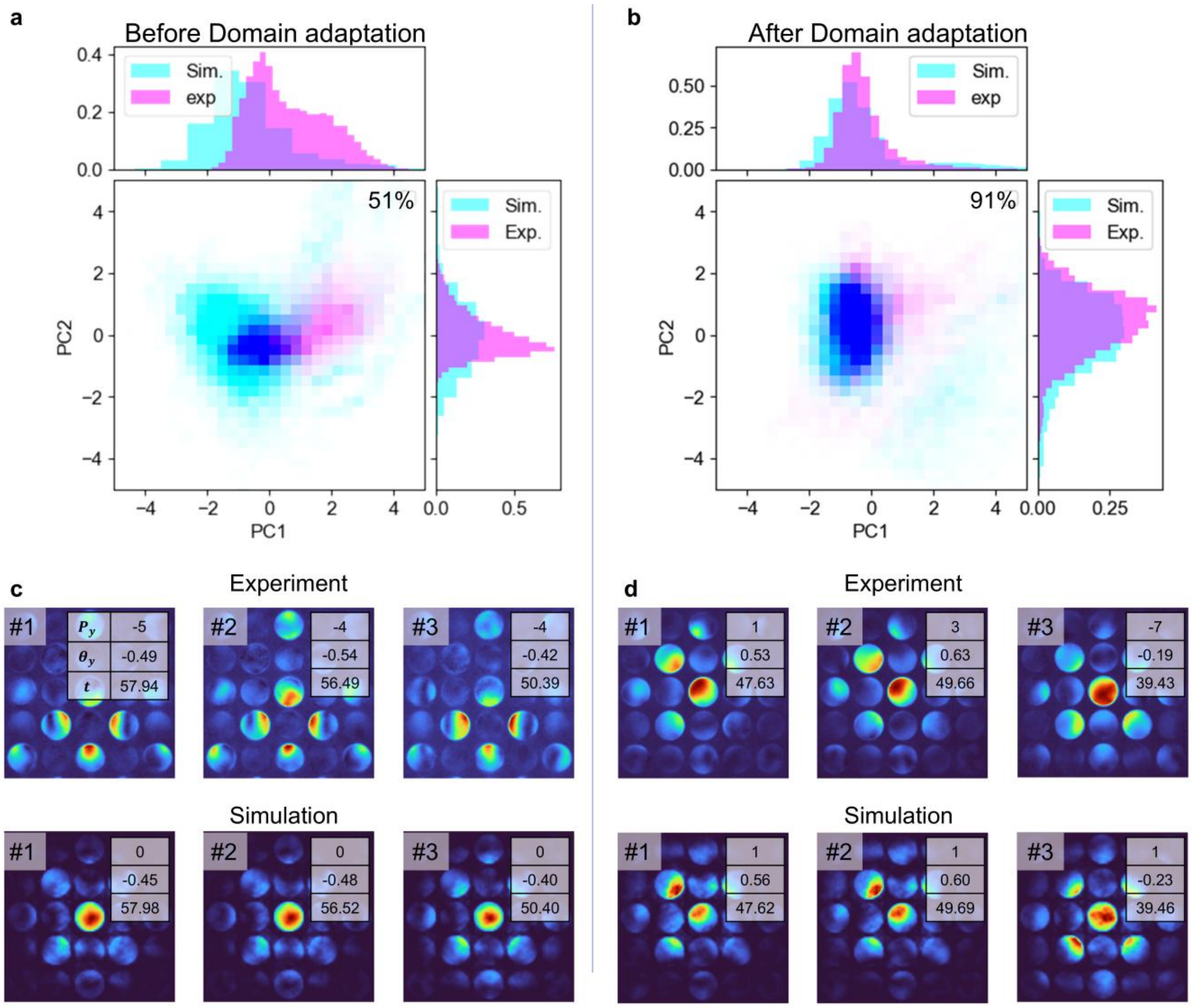

**Extended Data Fig. 1 | Evaluation of the effectiveness of domain adaptation on experimental 4D-STEM data. a, b**, 2D maps of latent vectors in PCA space represented by the two principal components (PC1 and PC2) for simulated (cyan) and experimental (magenta) datasets, accompanied by histograms showing the distributions of PC1 and PC2, before (**a**) and after domain adaptation (DA) (**b**). **c**, **d**, Exemplary DPs and the corresponding representative parameters ($P_y$, $\theta_y$, $t$) determined before (**c**) and after DA (**d**). The unit of parameter is: μC/cm$^2$ for $P_y$; mrad for $\theta_y$; nm for $t$. In each figure set, the upper and lower panels are experimental DPs and simulated DPs, respectively, which are closely aligned in the PCA space. After the DA, image similarity and accuracy in parameter prediction are enhanced dramatically.

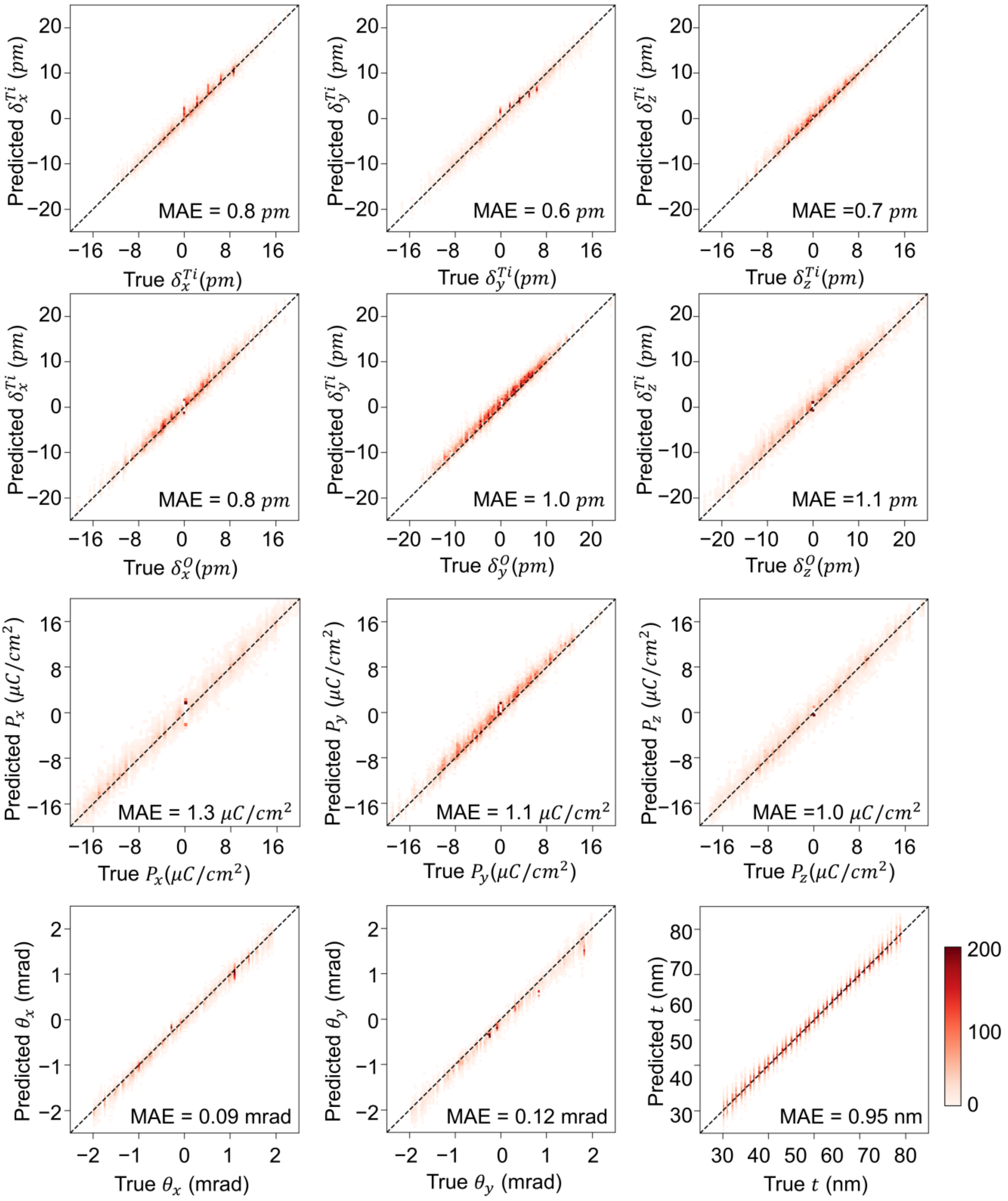

**Extended Data Fig. 2 | Evaluation of the machine learning model performance**. Parity plots comparing the actual property values with the values predicted by the machine learning (ML) model. All data points shown were tested on an out-of-sample dataset. The color bar represents the density of data points in each bin. The dashed line indicates perfect agreement (parity) between predicted and actual values. The mean absolute error (MAE) is provided for each plot.

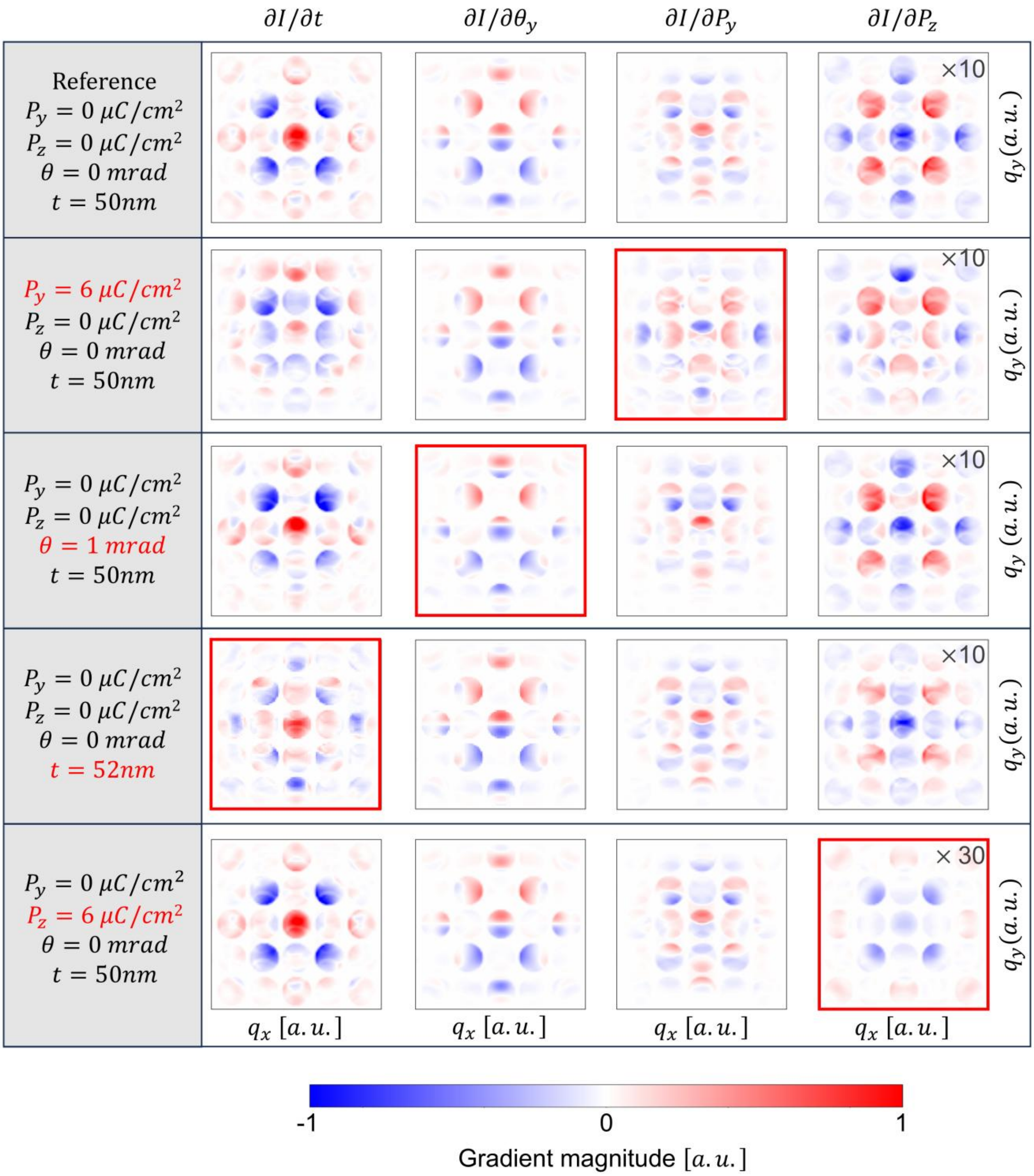

**Extended Data Fig. 3 | Derivatives of diffraction intensity with respect to sample thickness, tilt, and polarization components, demonstrating the independent identifiability of each variable.**

Each image represents the change in diffraction intensity, $\partial I/\partial x = I_{p+\delta p} - I_{p,\text{ base}}$, where $I_{p,\text{ base}}$ is the calculated intensity under the baseline conditions specified by labels on the left side, and $I_{p+\delta p}$ is the intensity calculated after infinitesimally increasing the parameter $x$ indicated by the topmost label. Relative to the baseline conditions in the first row, images in subsequent rows illustrate changes resulting specifically from altering the parameter highlighted in red. Images corresponding to the varied parameter exhibit noticeable intensity changes, whereas other images remain unaffected.

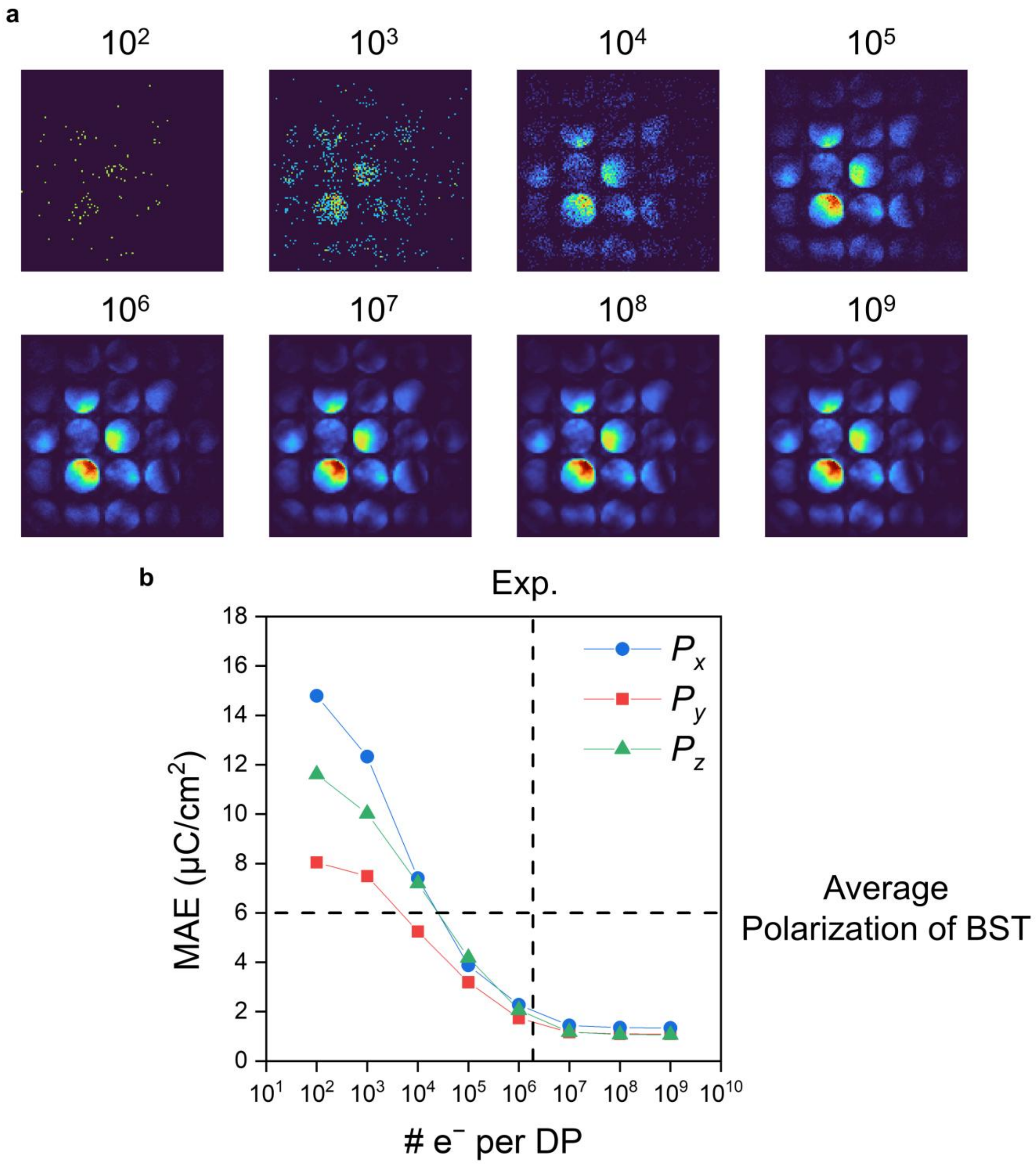

**Extended Data Fig. 4 | Electron dose required for detecting polarization signals in experimental 4D-STEM data. a,** 4D-STEM DPs with increasing total electron dose per DP. A total electron dose exceeding $10^4$ electrons per DP is necessary to reliably detect intensity variations between diffraction disks and intensity distributions within individual disks. **b,** Quantitative analysis showing the mean absolute error (MAE) in measuring polarization components as a function of total electrons per DP. The vertical dashed line marks the average dose level in our experimental DPs. The horizontal dashed line indicates the typical polarization magnitude observed in BST.

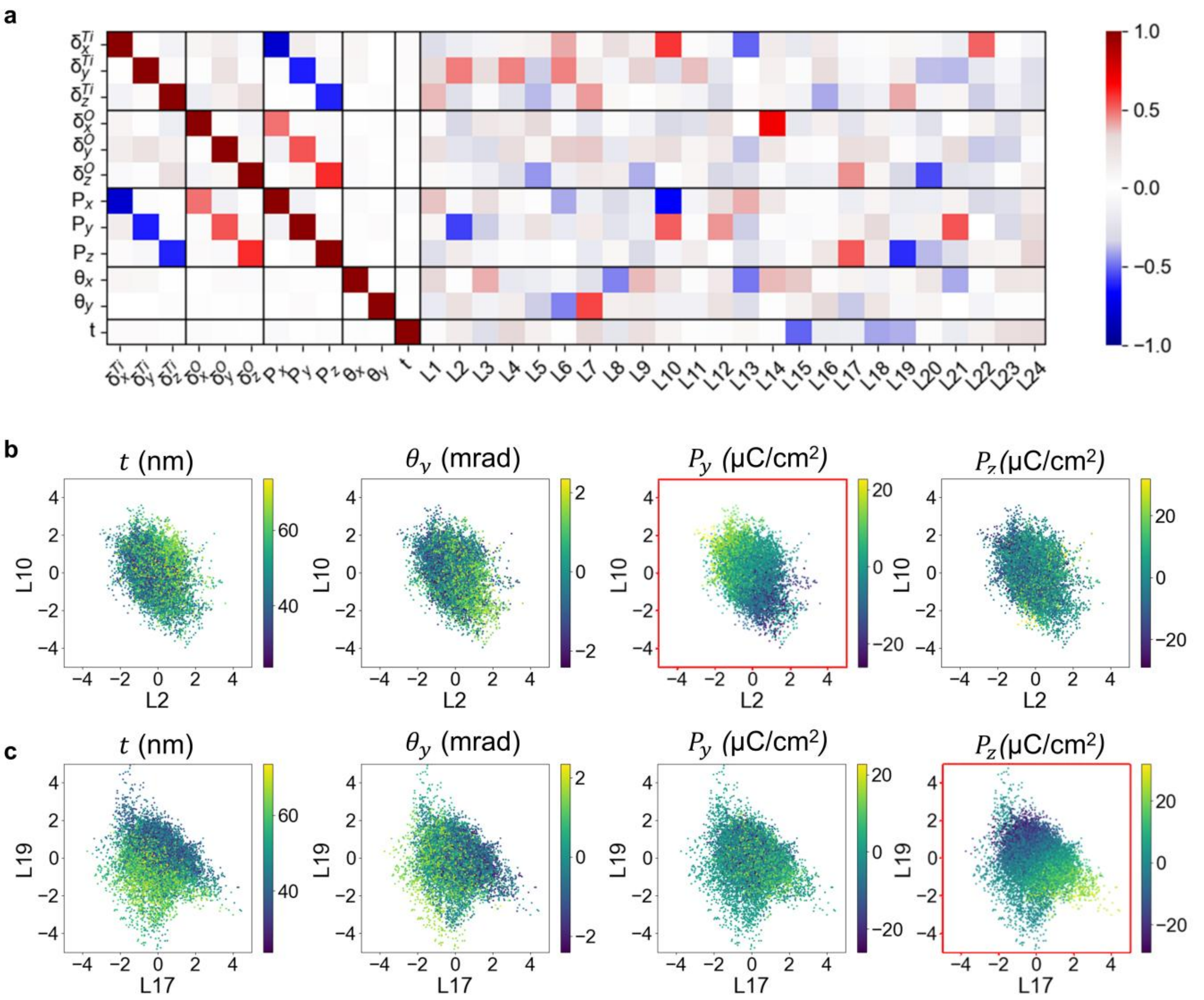

**Extended Data Fig. 5 | Pearson correlation analysis between predicted physical properties and latent vectors. a,** Heatmap of Pearson correlation coefficients depicting relationships among physical properties (left block) and between these properties and latent vectors (right block). Red indicates positive correlations, and blue indicates negative correlations. **b,c,** Scatter plots displaying data projections onto selected 2D latent subspaces ($L2$ vs. $L10$ in **b** and $L17$ vs. $L19$ in **c**). Data points are colored according to the value of different physical properties ($t$, $\theta_y$, $P_y$, and $P_z$, indicated by titles), highlighting their representation in latent space. Plot for $P_y$ in panel (**b**) and $P_z$ in panel (**c**), outlined in red, emphasize the strong alignment of these parameters within the respective latent subspaces.

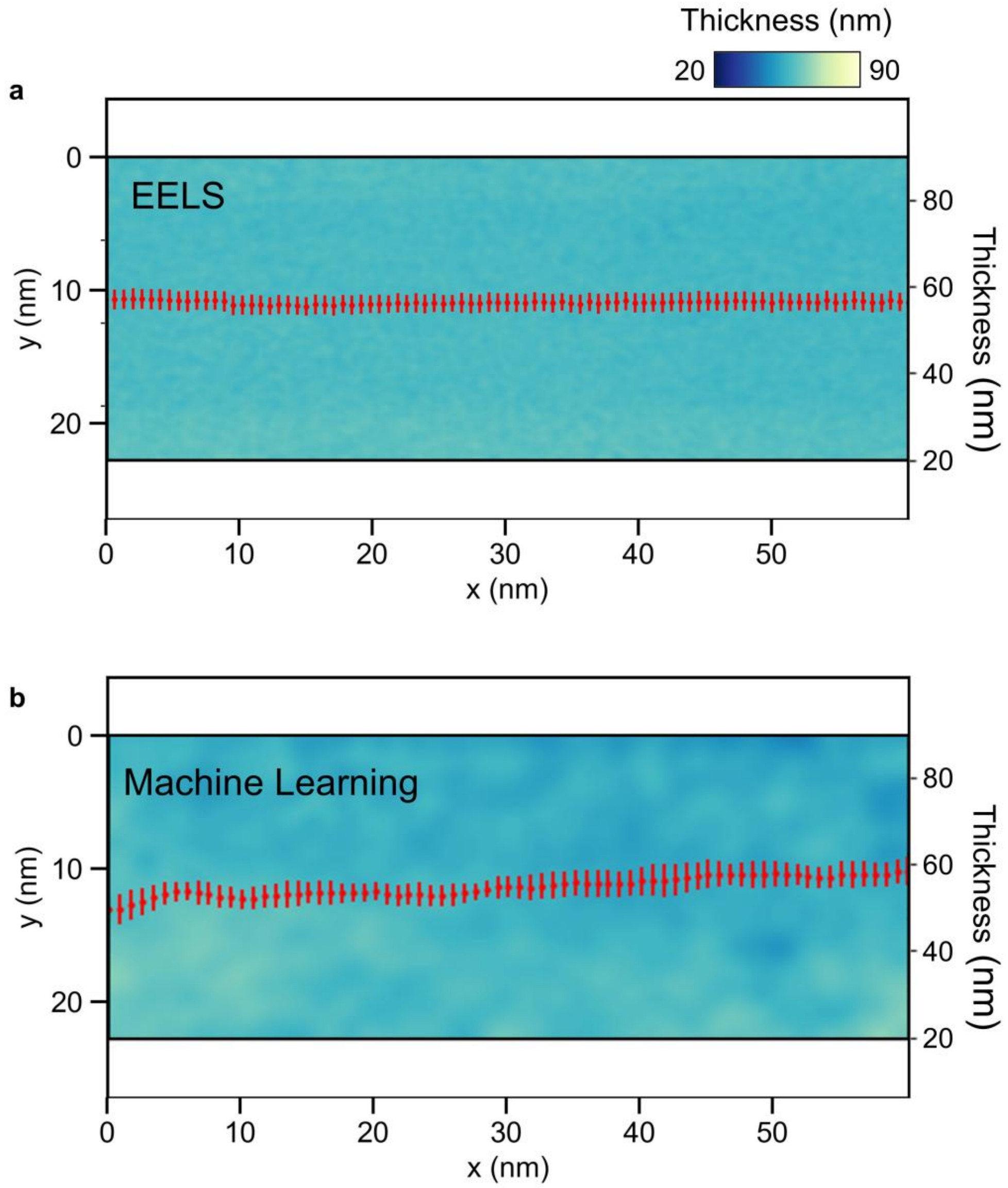

**Extended Data Fig. 6 | Thickness maps of TEM sample obtained by EELS and deep learning of 4D-STEM data. a,** Thickness map ($t/\lambda$) obtained from the electron energy-loss spectroscopy (EELS) spectrum image by using the log-ratio method. The inelastic mean free path ($\lambda$) of 117 nm was used for BST. **b,** Thickness map of the same region of BST obtained by deep learning-based analysis of 4D-STEM data. Color bar indicates thickness in nm. Overlaid red bar graphs represent the thickness profile along the x-direction, showing the mean value and variation are in good agreement.

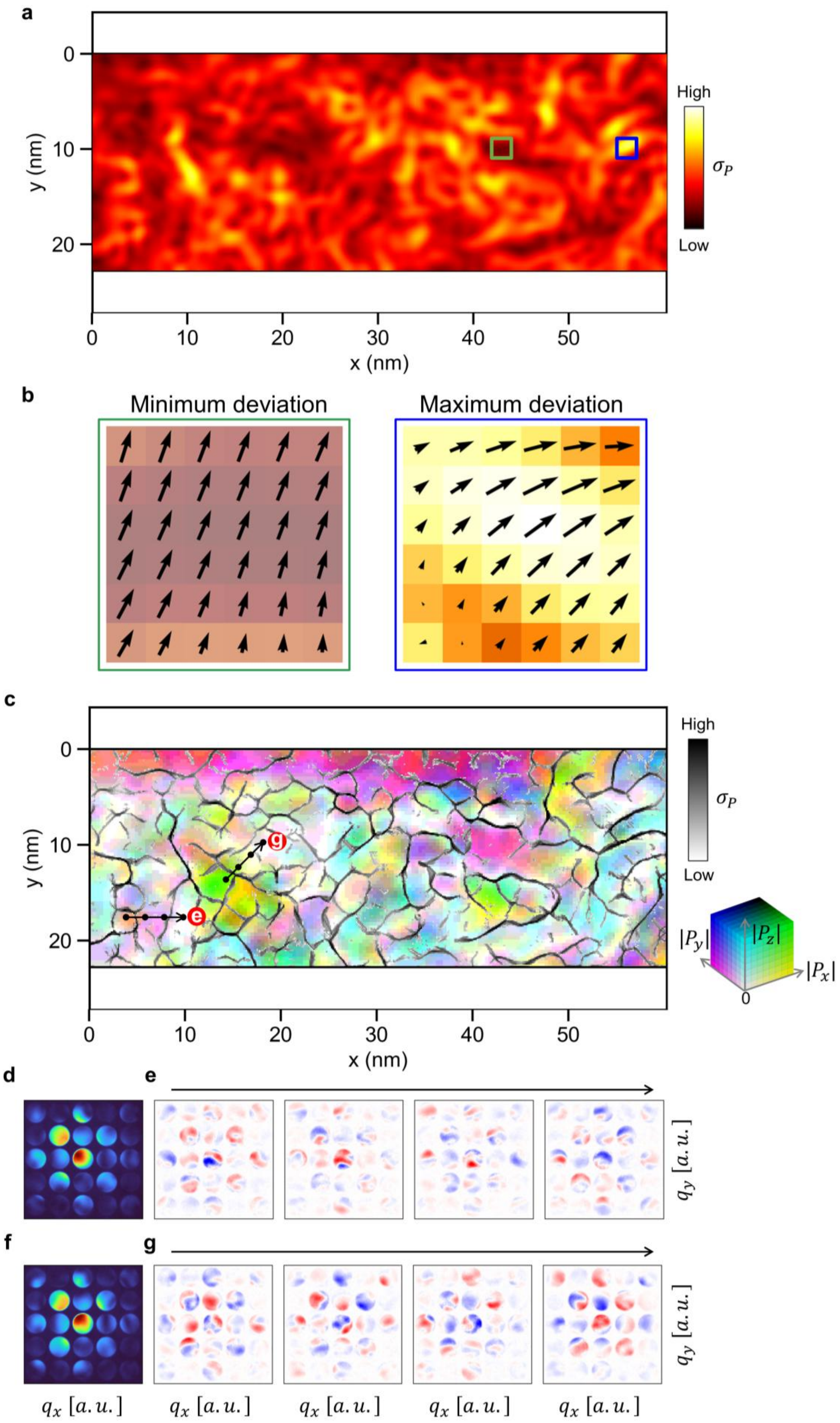

**Extended Data Fig. 7 | Domain structure characterized by defining domain boundaries. a**, Standard deviation ($\sigma_P$) map of polarization variation used to define domain boundaries (see Methods). The green and blue boxes highlight specific regions of interest detailed in **b**. **b,** Comparison of polarization vectors in a region of low variation (left, from green box in **a**) and high variation (right, from blue box in **a**). **c**, Domain boundaries (gray lines) obtained via skeletonization of $\sigma_P$ map overlaid on the polarization distribution map. The domain boundaries are displayed in gray scale to represent different extents of polarization variation. Variation of DPs from the points marked along the path (**e**) and (**g**) are analyzed in (**d-g**). **d,f**, Averaged DPs from the four data points along the path (**e**) and (**g**) in **c**, respectively. **e,g**, A series of the derivative DP patterns at each data point with respect to the average DPs in **d** and **f,** respectively. Slight but distinct variations of DP across the domain boundaries are discernible.

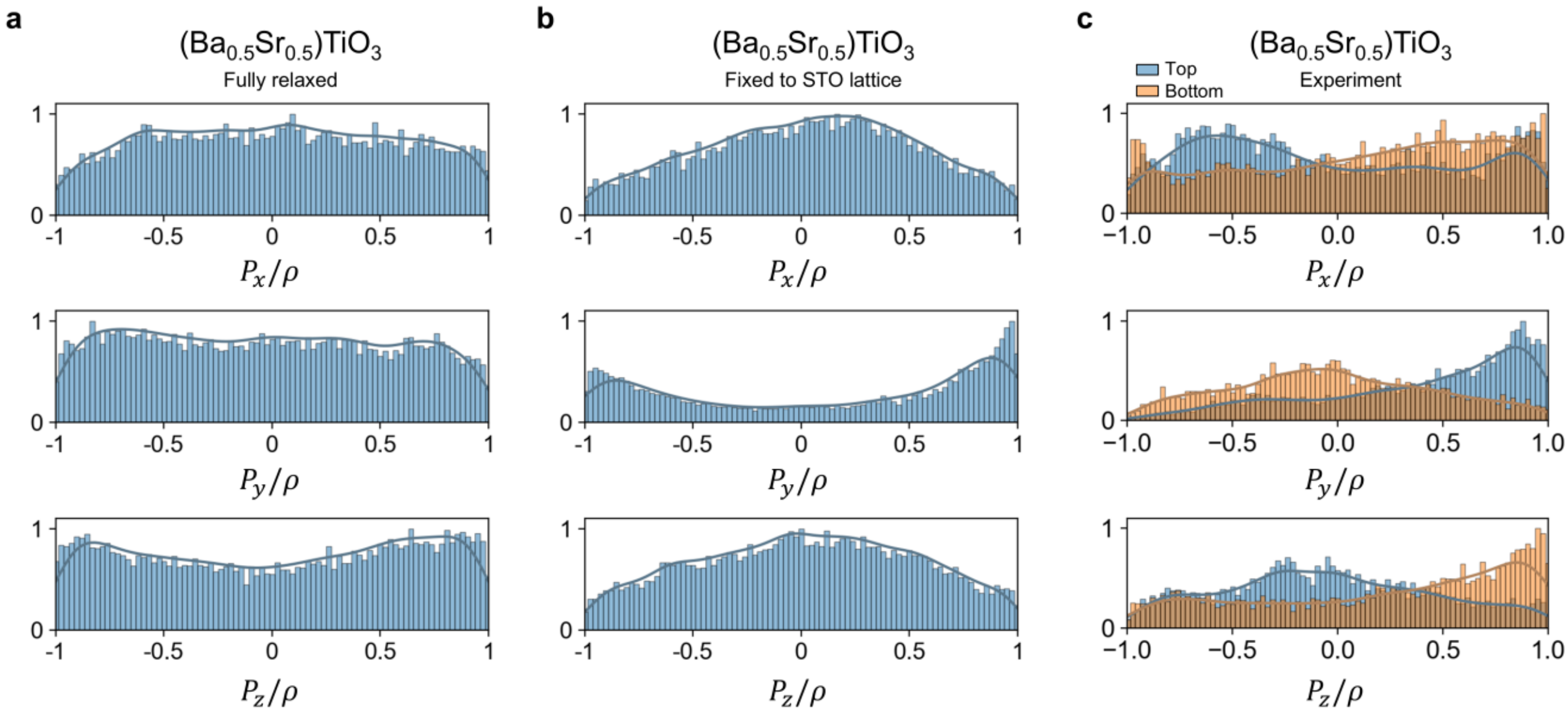

**Extended Data Fig. 8 | Effects of strain on the tail-to-tail polarization configuration of epitaxial BST film. a-c,** Histograms showing the distribution of normalized polarization components ($P_x/\rho$, $P_y/\rho$, $P_z/\rho$) in BST under different strain conditions: **a,** fully relaxed BST (MD), **b,** compressively strained BST to fit to the STO lattice (MD), and **c,** experimental data from the top and bottom regions of the BST film. Lines represent fitted probability density functions.

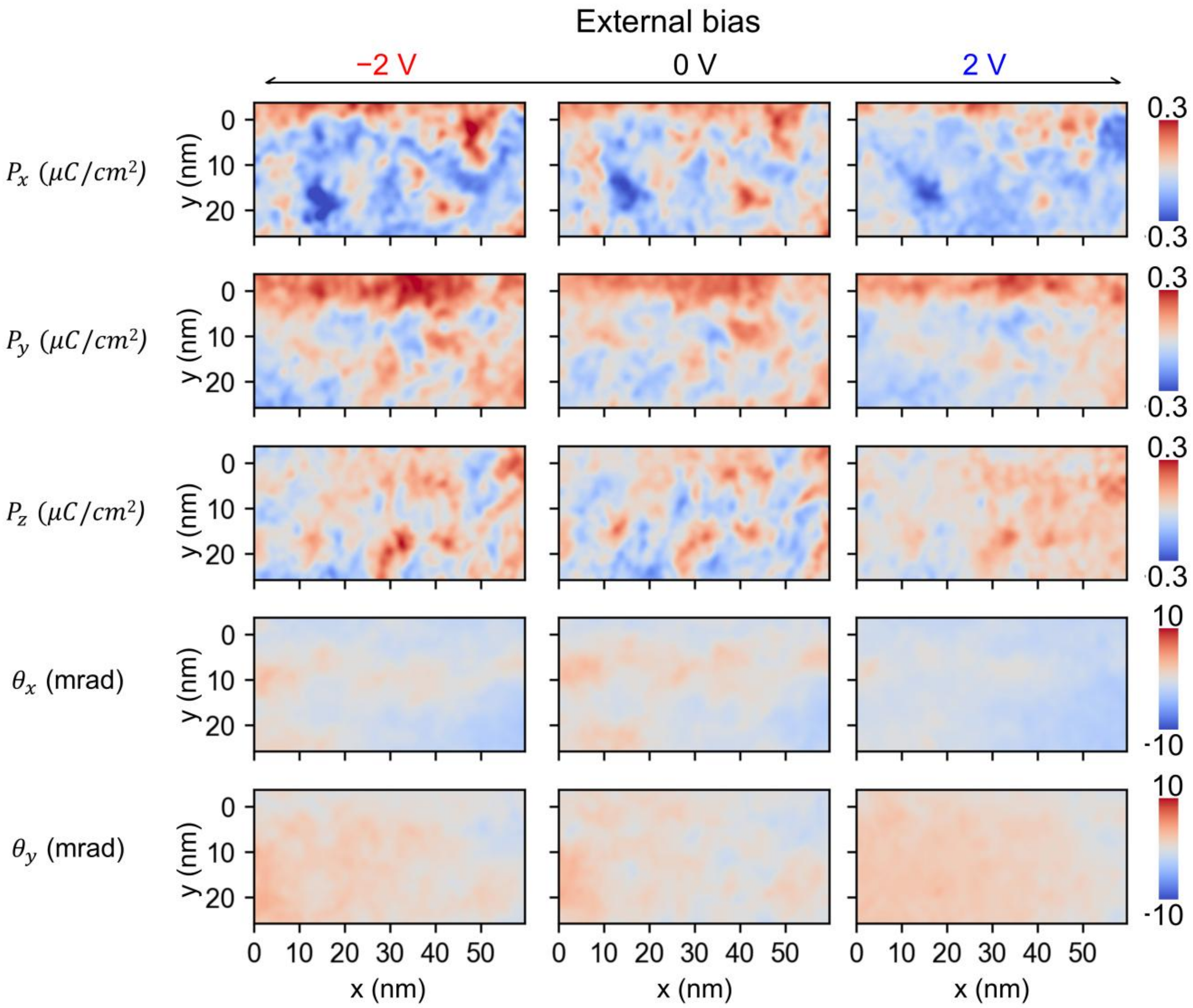

**Extended Data Fig. 9 | Verification of deep learning-based 4D-STEM analysis under external bias: reactive (polarization) vs. unreactive (tilt) components.** Polarization $(P_x, P_y, P_z)$ and tilt $(\theta_x, \theta_y)$ component maps obtained at -2 V, 0 V, and +2 V from deep learning analysis of 4D-STEM data. Each row shows one of the five components $P_x, P_y, P_z$, $\theta_x, \theta_y$, and each column corresponds to a different bias condition. The polarization maps change with applied bias, whereas the tilt maps remain unchanged, confirming that the method correctly isolates the reactive (polarization) from the unreactive (tilt) contributions.

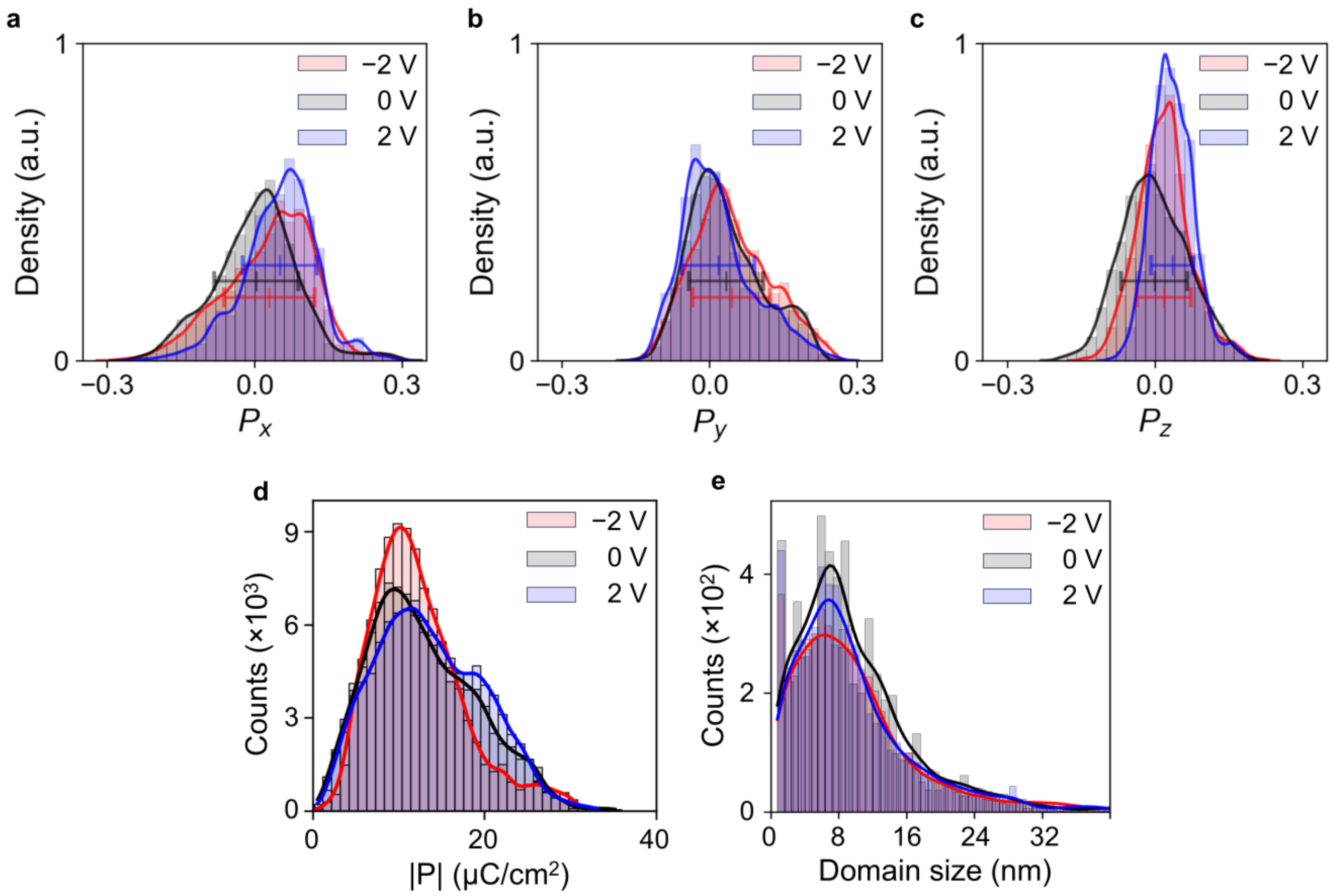

**Extended Data Fig. 10 | Statistical analysis of polarization vectors under electric fields. a, b, c,** Probability density distributions of the individual polarization components, $P_x$ (**a**), $P_y$(**b**), and $P_z$(**c**), at external biases of -2 V (red), 0 V (black), and +2 V (blue). **d,** Histograms of the polarization magnitude (|P|) under the same bias conditions. **e,** Histograms of domain sizes at -2 V, 0 V, and +2 V.